\documentstyle[editedvolume,epsfig]{crckapb}

\begin{opening}
\title{A Review of High-Redshift Merger Observations}
\author{R. G. Abraham} 
\institute{Royal Greenwich Observatory\\ 
Madingley Road, Cambridge\\ CB3 OEZ United Kingdom}

\end{opening}

\runningtitle{Review of High-Redshift Mergers}

\begin{document}

\begin{abstract}
Evolution in the merger rate as a function of redshift is {\em in principle} the
key observable testing hierarchical models for
the formation and evolution of galaxies. However, {\em in practice},
direct measurement of this quantity has proven difficult. In this
opening review I outline the current best estimates for the merger rate as a
function of cosmic epoch, focussing mostly upon recent advances made
possible by deep ground-based redshift surveys and morphological
studies undertaken with HST. I argue that a marriage of these
techniques, in an attempt to determine the space density of mergers
amongst the abundant morphologically peculiar population at high
redshifts, is probably the most promising currently-available avenue
for determining the prevalence of mergers at high redshifts.  However,
resolved kinematical studies, which seem set to become available in
the next few years, are probably the best hope for a definitive
determination of the space density of mergers at high redshifts.
\end{abstract}

\section{Observational Techniques:}

Three observational techniques have been used to probe changes in the
merger rate as a function of cosmic epoch. These are (a) studies of
angular and physical correlation functions, (b) pair counts, and (c)
morphological studies. The advantages ($+$) and disadvantages ($-$) of
each of these approaches for studying high-redshift mergers can be crudely summarized as follows.

\subsection{Correlation Functions }
\begin{list}{$+$}{}
\item A close connection to large-scale structure work via the clustering
statistics $w(\theta)$ and $\xi(r)$. Since mergers can be identified
as the end-products of large-scale clustering, changes in the
correlation function at small radius provide a conceptual link between
the large-scale and small-scale regimes being probed with the
same statistics. 
\item Biases inherent in measuring correlation functions are
fairly well-understood. 
\end{list}
\begin{list}{$-$}{}
\item Measurement of correlation functions is best suited to large samples.
\item These statistics are hard to measure at small radii,
particularly when galactic components (such as giant HII regions)
become difficult to distinguish from merging companions.  This is
often the case when probing distant galaxies in the rest-frame
ultraviolet, {\it eg.} in the Hubble Deep Field, where knots of
star-formation cannot easily be distinguished from companions (Colley
et al. 1996).
\item {\em Not really measuring merger rate (see below)}
\end{list}

\subsection{Pair Counts Approach}
\begin{list}{$+$}{}
\item Simple statistics to measure (described in detail in the next section).
\item Conceptually the pair-counts approach is an integration over the
two-point correlation function, with better signal-to-noise properties
at small radii, so the biases are also fairly well-understood.
\end{list}
\begin{list}{$-$}{}
\item Like correlation functions, the statistic becomes ambiguous when
merging companions become indistinguishable from galactic components.
\item {\em Not really measuring merger rate (see below)}
\end{list}

\hspace{0.5cm}

\noindent The most important criticism common to both the correlation
function and pair counts approaches is that neither provides a direct
measure of the merger rate. The fundamental difficulty is {\em the
uncertain physical timescale over which merging occurs}, given
physical proximity between galaxies. Both statistics probe the fuel
``reservoir'' available for close gravitational interaction, but what
we really seek is an understanding of the rate at which the merger
``engine'' operates in converting these galaxies into the by-products
of mergers. To understand this, one must observe the mergers in
progress.  So the third approach must necessarily be a morphological
one:

\subsection{Morphological Approach}
\begin{list}{$+$}{}
\item Direct observation of {\em mergers in progress}
\end{list}
\begin{list}{$-$}{}
\item Poorly-understood biases ({\em eg.} morphological K-corrections)
\end{list}

Because of space limitations, and because they are well-reviewed
elsewhere, this review ignores studies of the
correlation function (eg. Neuschaefer et al. 1997), and touches only
rather superficially on recent studies on pair-counts in \S2, in order
to focus mostly on summarizing very recent progress made on
morphological studies at high-redshift with HST in \S3.

\section{Evolution in Density of Pairs}

Because existing redshift surveys are not yet deep enough to allow a
direct analysis of pair-density evolution to be undertaken, pair count
studies can be broadly grouped into two categories: (1) pure
photometric analyses with no redshift information, and (2) photometric
searches around galaxies with known redshifts.  A flavor for the
methodology adopted in the latter ``mixed'' category of
redshift-photometric surveys is sketched out in Figure 1, which
illustrates the technique adopted by Yee \& Ellingson (1995). Searches
are conducted within an aperture projected onto the sky, defined by a projected {\em physical} radius of $R_1 = 20 h^{-1}$
kpc of a galaxy of known redshift. But because low-redshift galaxies will
have a large search radius that is likely to be littered with faint
unrelated background galaxies, additional restrictions must be placed
on the search radius based on the apparent magnitude of the putative
companion galaxy. Each companion must lie within a radius $R_2(m) =
D_{nm}(m)/3$ where $D_{nm}(m)$ is the expected nearest neighbor
distance for galaxies brighter than companion magnitude
$m$. ($D_{nm}(m) = {1 \over {2\sigma(m)^{1/2}}}$ (Rose 1977), where
$\sigma(m)$ is the surface density of galaxies brighter than $m$.)  A
correction can then be made for the expected mean number of unrelated
galaxies in the search area, assuming Poisson statistics:
$\overline{N} = P_o\left[ 1 + 2\ln {R_1\over R_2(m_l)} \right]$

\begin{figure}
\parbox{2.5in}{
\epsfig{figure=ring.ps,width=2.5in}
}
\parbox{2in}{
    \footnotesize
    \hbox{\em Figure~1.} Cartoon showing the search criteria for the ``mixed''
    redshift-photometric analysis of pair-density evolution undertaken by
    Yee \& Ellingson (1995). The search radius $R_1$ is fixed by the physical
    scale being probed, while radius $R_2$ varies with the companion magnitude. 
}
\end{figure}
\setcounter{figure}{1}

Results from all published surveys of pair-density evolution (from both
classes of survey) are summarized below, in the form of power-law
evolution in the density of pairs $(1 + z)^n$:

\subsection{Pure photometric analyses}
\begin{enumerate}
\item Zepf \& Koo (1989) $\Rightarrow (1+z)^{4.0\pm2.0}$
\item Burkey et al. (1994) $\Rightarrow (1+z)^{3.5\pm0.5}$
\item Carlberg, Pritchet, \& Infante (1994) $\Rightarrow (1+z)^{3.5\pm1.0}$
\item Woods, Fahlman, \& Richer (1995) $\Rightarrow$ No evolution!
\end{enumerate}

\subsection{Searches around galaxies with known redshifts}
\begin{enumerate}
\item Yee \& Ellingson (1995) $\Rightarrow (1+z)^{4.0\pm1.5}$
\item Patton et al. (1997) $\Rightarrow (1+z)^{2.8\pm0.9}$
\end{enumerate}

The recent work by Patton {\it et al.} is based on the largest
redshift sample to date: 545 field galaxies with a mean redshift of $z
= 0.3$.  Clearly these published surveys do not yet probe very far out
in redshift space, but it is curious that with the exception of the
pure photometric study by Woods and collaborators, all pair-count work to date (but not all correlation function work, eg. Neuschaefer et al. 1997) is
roughly consistent with the $(1+z)^{2.7\pm0.5}$ increase in co-moving
luminosity density from the Canada-France Redshift survey
(Lilly~et~al.~1996). It is perhaps worth re-stating that the pair
fraction increase is {\em not} the same thing as the merger rate
increase: conversion between these is sensitive to assumptions made
with regard to merger timescales. No particular consensus exists
amongst the various authors regarding the appropriate conversion
between pair density and merger rate.  For example, given pair
evolution with the form $(1+z)^n$, the merger rate is variously
assumed to take the form $(1+z)^{n-1}$ by Burkey et al., $(1+z)^{n+1}$
by Carlberg et al., and $(1+z)^n$ by Yee \& Ellingson.  It is
interesting that from the distribution of the projected distances of
companions as a function of redshift, Patton {\it et al.}  conclude
that mergers are likely to occur over timescales of 150~Myr --
400~Myr.

\section{High-Redshift Morphology}

Convincing observations of high-redshift galaxy morphology have only
become possible recently, with WF/PC2 observations on HST, and the
resulting flood of imaging data has resulted in a great increase in
our understanding of the field galaxy population at high redshifts.
Selected highlights (with apologies to many colleagues whose work cannot be included here due to space limitations) from recent HST-based imaging work, relevant to
high-redshift mergers, are summarized below. I then go on to describe
what seem to me a number of ``key issues'' with regard to
understanding the connection between these imaging observations and
high-redshift merger scenarios.

\subsection{HST Field Surveys with Morphological Information}

Early work from the Medium Deep Survey (MDS) reported an excess of faint
peculiar systems (Griffiths~et~al.~1994), but the excess become much more
convincing with the extension of this work to include number counts as
a function of magnitude (Glazebrook~et~al.~1995; Driver~et~al.~1995; Abraham~et~al~1996). Worries with regard to bulk misclassification of peculiar galaxies have been eased by the incorporation of objective machine-based classifications, which are now routinely being used to supplement (and in some cases replace) visual morphological classifications from a number of different surveys (Abraham~et~al.~1996; Odewahan~et~al.~1996; Naim~et~al.~1997; Brinchmann et al. 1997). The incorporation of redshift information, probing the regime out to roughly $z<1$, has generally confirmed the photometric work, and in turn pushed much further in terms of our understanding of the contributions of the morphological classes to the star-formation history of the Universe (Cowie,~Hu,~\&~Songaila~1995; Schade et al. 1996; Pascarelle et al. 1996; Brinchmann et al. 1997). Imaging follow-up observations of Lyman-limit systems  have in turn yielded views of systems at $z>2$ (Giavalisco et al. 1996). Because of their unprecedented depth, and intensive spectroscopic and photometric redshift follow-ups (Cohen et al. 1996; Lanzetta et al. 1996; Mobasher et al. 1996; Lowenthal et al. 1997; Phillips et al. 1997; Sawicki et al. 1997), Hubble Deep Field observation of faint galaxy morphology  are the likely to yield the greatest insights into evolving distribution of morphological types at high redshifts for some time to come (Abraham et al. 1996; van den Bergh et al. 1996; Bouwens et al. 1997; Guzman et al. 1997). Eagerly anticipated are the results from high-resolution infrared observations in the HDF with NICMOS, and forthcoming observations of the HDF South. Deeper observations (and, perhaps more importantly, greater sky coverage) will perhaps have to await completion of the HST Advanced Camera, and hopefully NGST.

While recent work has left little doubt that much of the high-redshift Universe is morphologically peculiar (about 1/3 of all galaxies by $I_{300W}\sim24$ mag), the nature of these systems is currently unknown. In the context of the present meeting, the bottom-line question seems to be: what fraction of morphologically peculiar galaxies are mergers in progress? Unfortunately this question
cannot be answered at the present time. Part of the problem is simply
one of subjective and confusing taxonomy used to describe the morphologies of faint galaxies. For example, are ``Morphologically Peculiar'' (Griffiths et al. 1994), ``Chainlike'' (Cowie, Hu, \& Songaila 1995), ``Blue Nucleated'' (Schade et al. 1996), ``Irregular/Peculiar/Merger'' (Glazebrook et al. 1995; Abraham et al. 1996a,b; Brinchmann et al 1997), and ``Tadpole'' (van den Bergh et al. 1996) galaxies similar objects simply denoted by different names, or a true reflection in the diversity of the morphologies of galaxies in the distant Universe? At a more mundane level, is bandshifting of the rest-frame of observation (the so-called ``morphological K-correction'') playing an important role, by fooling us into mistaking systems with normal UV morphology for new classes of galaxy? I have argued (Abraham et al. 1996a;b) that problems with both taxonomy and morphological K-corrections can be circumvented by abandoning conventional galaxy classification schemes in favor of objective classifications, based on measurement of simple structural parameters (such central concentration and asymmetry), and by calibrating such measurements against simulations of the appearance of galaxies at a range of redshifts. The best example of this approach is recent work by Brinchmann et al. (1997). In this work, HST imaging and ground-based spectroscopic data for $\sim 300$ galaxies with good completeness to $I<22$ mag are used demonstrate that the peculiar galaxy excess is already in place at redshifts where morphological K-corrections have a minimal effect. Simple and objective approaches to quantifying galaxy morphology seems to me the safest course until the observational biases are better understood, but of course one pays a price by ``smoothing over'' much of the interesting diversity (denoted with correspondingly charming nomenclature) seen in galaxy forms on deep images.

\begin{figure}
  \begin{center}
  \epsfig{figure=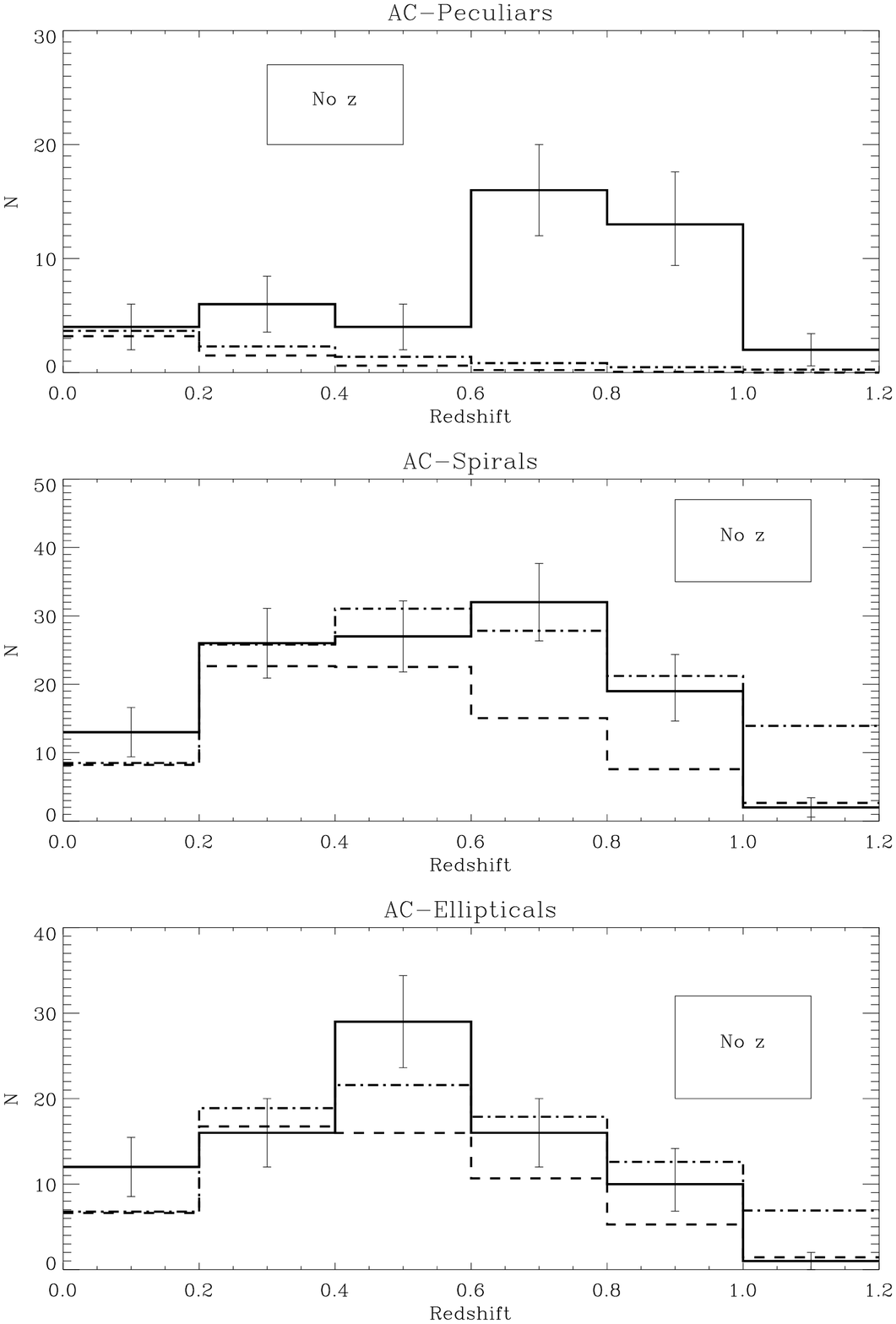,height=4.5in,angle=0}
  \end{center}
\caption{Morphologically segregated redshift distributions (taken from Brinchmann et al. 1997) for irregular/peculiar/merging galaxies (top), spirals (middle), and ellipticals (bottom). The sample consists of $\sim 300$ galaxies from the CFRS (Lilly et al. 1995) and LDSS (Ellis et al. 1996) redshift surveys. Classifications have been based upon measurements of central concentration and asymmetry on deep HST WF/PC2 $I_{814}$-band images, calibrated
by simulations for the effects of bandshifting of the rest-frame of observation. (Although over the redshift range probed such effects are small.) Dashed and dot-dashed lines are the predictions of no-evolution, and mild evolution models. Note the marked excess in the number of irregular/peculiar/merging systems at high redshifts.}
\end{figure}

What fraction of this diverse set of peculiar galaxies are actually mergers in progress? The problems inherent in answering this question
are illustrated in Figure 3, which shows candidate Lyman limit systems in the Hubble Deep Field with $I_{814} < 25$ mag. Obviously most are morphologically peculiar, but how many readers can honestly claim that these resemble the appearance of local merging systems? The difficulties inherent in identifying mergers amongst the distant morphologically peculiar population are made clear by simulations recently published by Hibbard and Vacca (1997), who use HST WF/PC camera ultraviolet data to predict the appearance of the high-redshift counterparts to local merging starburst systems. Because the usual signatures of mergers (tidal tails, distorted disks) are no longer visible at high redshifts, merging starbursts seem to provide at least qualitatively reasonable counterparts to many faint peculiar galaxies (Fig.~4). While very suggestive, such simulations need to be interpreted with some caution, because they do not explicitly account for evolutionary effects which are known to be important at high redshifts. For example, in the simulations shown in Fig. 4 the smooth components of the galaxy are invisible largely because their stellar populations are evolved in the local reference images and are subject to strong K-corrections. However at $z\sim 2$ {\em all} stellar populations are relatively young. More detailed simulations incorporating explicitly the effects of evolution would be valuable.

\begin{figure}
  \begin{center}
  \end{center}
\caption{Candidate $U_{300}$-band dropout Lyman-limit systems in the Hubble Deep Field,  with $I<25$ mag, taken from van den Bergh et al. (1996). ``True-color'' images were constructed by
combining the $U_{300}$,$V_{606}$, and $I_{814}$-band observations.}
\end{figure}

\begin{figure}
  \begin{center}
  \end{center}
\caption{(Top montage:) Simulations (based on ultraviolet HST WF/PC  observations) taken from Hibbard \& Vacca (1997), showing the predicted appearance of local well-known merging starburst galaxies as seen at high-redshifts in the Hubble Deep Field. (Bottom montage:) Morphologically similar galaxies in the Hubble Deep Field.}
\end{figure}

Clearly the best way forward will be to incorporate dynamical information to determine directly which peculiar galaxies show distinct kinematical sub-components. Unfortunately these observations are not currently feasible, although they may soon become possible with adaptive optics and the new generation of 8m-class telescopes. In the meantime, a promising approach to quantifying the fraction of mergers amongst the distant peculiar galaxy population may be to measure statistics which are relatively insensitive to image distortions resulting from bandshifting and surface-brightness biases, but which track probable merger activity. One such statistic is the ``Lee Ratio'', a measure of image bimodality. This statistic been applied to images of galaxies in the CFRS survey (Fig. 5) and to HDF galaxies, with the result that around $\sim 40\%$ of faint peculiar systems are significantly bimodal, with an $\sim (1+z)^3$ increase in the merger rate (Le F\'evre et al., in preparation). 

\begin{figure}
  \begin{center}
  \epsfig{figure=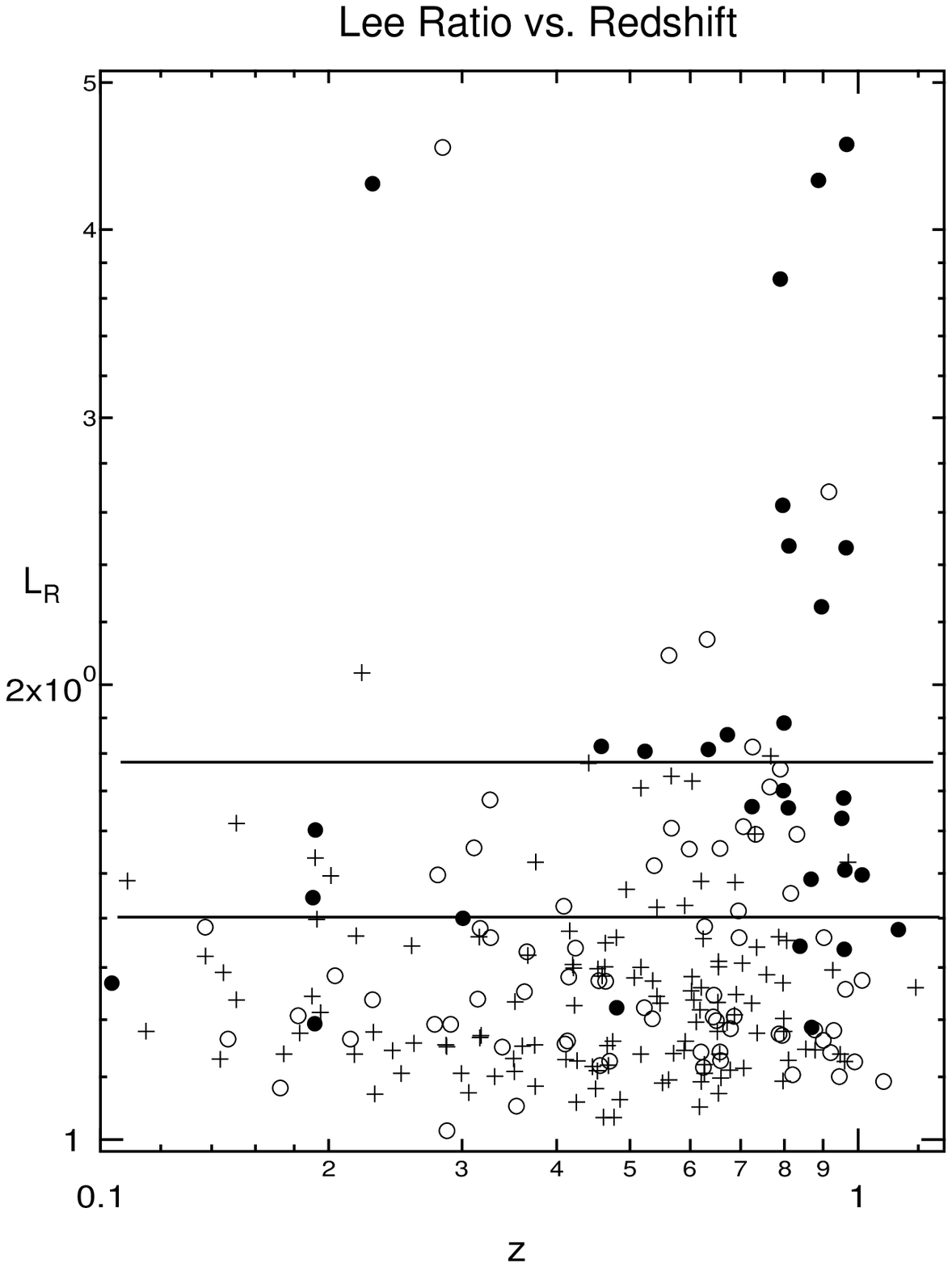,width=3.5in,angle=0}
  \end{center}
\caption{Lee ratio $L_R$ bimodality index for HST WF/PC2 images of galaxies in the Canada-France Redshift survey, taken from Le F\'evre et al. (in preparation). Plot symbols correspond to visual classifications of the galaxies as ``major mergers'' (solid circles), ``minor mergers'' (open circles), and undisturbed galaxies (crosses). There is a marked increase in the number of merger candidates at $z>0.5$. Note also the fairly good agreement between classification as a major merger and high Lee ratio.}
\end{figure}

Another promising approach to quantifying the fraction of high-redshift mergers is to better understand the nature of the star-formation activity in distant peculiar galaxies, in order to test if their star-formation histories are consistent with merging. One overlooked method for accomplishing this is modelling of the {\em internal} pixel-by-pixel colors of these galaxies. This approach splits the galaxy up into components under the assumption that morphology can be used to identify stellar populations. In principle, this breaks much of the degeneracy inherent in population synthesis modelling which treat galaxies as point sources, and puts constraints not only on merger activity, but also on the dust content, relative ages of disk and bulge, and general star-formation history. As an example, Figure~5 shows two ``chain galaxies'' in the Hubble Deep Field. The resolved color analysis indicates that these systems are showing well-organized star-formation activity -- a strong constraint on possible merging scenarios. The ``knots'' in these galaxies are unlikely to be individual galaxies;  star-formation seems to be triggered linearly along the body of the galaxy, starting from a initial ``seed'' starburst, with individual knots that are about as luminous as ``super star clusters'' that are the putative progenitors of globular clusters. If such systems are mergers in progress then a mechanism for igniting the chain reaction via mergers must be invoked. Work is currently in progress (Abraham et al. 1998) to apply these ideas to all peculiar galaxies in the HDF in order to determine which galaxies are the strongest merger candidates. Until kinematical studies become available with resolution sufficient to detect merging subsystems at high redshifts, color-based studies may be the best way to detect distinct physical sub-components in morphologically peculiar galaxies.

\begin{figure}
  \begin{center}
  \epsfig{figure=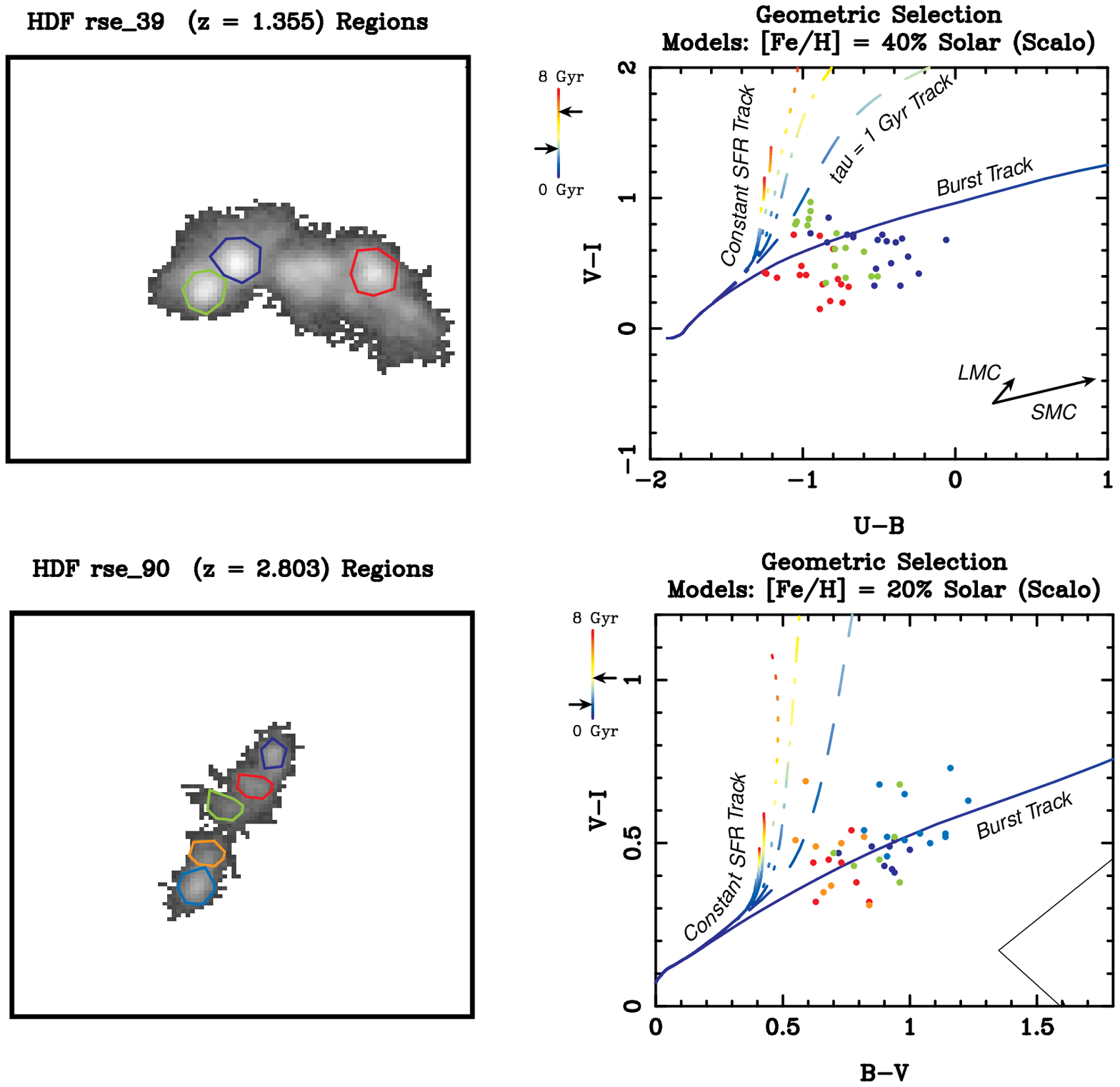,width=3.5in,angle=0}
  \end{center}
\caption{Figure 2: High redshift ``chain galaxies'' from the HDF (left), and corresponding pixel-by-pixel resolved color-color diagrams (right). Apertures on the images at left enclose subsets of pixels at right. Star-formation history tracks are shown as solid lines, keyed to the colored age bar. Note that both objects are consistent with pure protogalactic starburst tracks with ages $< 0.2$ Gyr. Synchronization in star-formation activity in the system at $z=1.355$ (with the ages of bursts changing monotonically with age from a seed knot of star-formation) is seen in many chainlike galaxies. Arrows on the age bar correspond to the age of the universe in an $\Omega=1$ and a low-$\Omega$ Universe. Arrows at the bottom-right corner of the right-hand panels are dust vectors. (Figure taken from Abraham et al., in preparation)}
\end{figure}

\section{Conclusion}

The most recent pair-count analyses seem to suggest that the fraction of physical pairs grows as $\sim (1+z)^3$. However, the conversion between pair fraction and merger rate is uncertain, and the pair count work so far published is limited to fairly low redshifts ($<z>\sim0.3$). At higher redshifts morphological work with HST indicates that by $I=24$ mag something over 30\% of all galaxies are morphologically peculiar. Simulations and follow-up spectroscopic work suggest this excess in morphologically peculiar systems is a physical effect, and not merely the result of ``morphological K-corrections''. The fraction of mergers amongst these morphologically peculiar galaxies is unknown, because obviously merging local systems, such as nearby major starburst galaxies,  no longer appear like conventional mergers at high redshifts. A preliminary analysis of image bimodality (a robust parameter that in principle flags major mergers even at high redshift) amongst $I<22$ mag peculiar systems with $z<1$ suggests that around 40\% the morphologically peculiar galaxies are strongly bimodal, and thus probably merging. This is consistent with a merger rate increases of $\sim (1+z)^3$. Internal color analyses of morphologically peculiar systems is an interesting  next step in probing high-redshift mergers, leading ultimately to kinematical investigations in the next few years.

\section{Acknowledgments}

I thank my collaborators Richard Ellis, Nial Tanvir, Sidney van den Bergh, Karl Glazebrook, Andy Fabian, and Basilio Santiago for their contributions to our high-redshift morphology projects. I also thank Jarle Brinchmann and Olivier 
Le F\'evre for permission to show figures in advance of publication, and Bill Vacca for useful discussions.

\end{document}